\documentclass[journal abbreviation, manuscript]{copernicus}
\nolinenumbers

\begin{document}

\title{Direct Lagrangian tracking simulation of droplet growth in vertically-developing turbulent cloud}


\Author[1][iwashima.m.6147@m.isct.ac.jp]{Masaya}{Iwashima} 
\Author[2][onishi.ryo@scrc.iir.isct.ac.jp]{Ryo}{Onishi}

\affil[1]{Engineering School, Department of Mechanical Engineering, Institute of Science Tokyo}
\affil[2]{Supercomputing Research Center, Institute of Integrated Research, Institute of Science Tokyo}




\runningtitle{TEXT}

\runningauthor{TEXT}

\received{}
\pubdiscuss{} 
\revised{}
\accepted{}
\published{}


\firstpage{1}

\maketitle

\begin{abstract}
  We developed a new explicit cloud microphysical model, based on direct numerical simulation (DNS) with Lagrangian particle tracking.
  The model employs a vertically-elongated quasi-1D computational domain extending from the ground to the cloud top to explicitly capture the vertical structure of clouds.
  This allows us to simulate the all warm-cloud microphysical processes, including activation, condensation growth, collision-coalescence growth, and sedimentation.
  A homogeneous isotropic turbulence field is incorporated into this domain to explicitly resolve the turbulent wind fluctuations.
  Cloud microphysics simulations with and without turbulent wind fluctuations were performed to clarify the impact of turbulence on droplet growth.
  We obtained new insights into the altitude- and time-dependent microphysical statistics, which cannot be obtained through conventional DNS researches for a cubic box domain with periodic boundaries.
  The comparison have shown that turbulence promoted the collision-coalescence growth of droplets.
  During the early developing stage, where the updraft was present, turbulence promoted the collisions between droplets with similar sizes (autoconversions) in the middle layer of the cloud.
  In later stage, relatively large droplets produced by autoconversions actively collected smaller droplets (accretions) in the middle and lower layers.
  The onset of precipitation at the ground occurred earlier and the first raindrop at the ground was larger in turbulence case than that in non-turbulence case.
\end{abstract}


\introduction  
In numerical simulations for atmospheric science, cloud modeling that represents cloud microphysics has been one of the central topics of discussion for more than a half century \citep{Grabowski_etal2019}.
Cloud modeling has generally adopted the Euler method, which treats grid-average quantities (e.g. mass or/and number of liquid water) as continuous fields.
The Euler method is generally classified into bulk method and spectral (bin) method.
In the bulk method, for several hydrometeors such as cloud water and rainwater, mass (or/and number) is/are used as prognostic variable(s) to calculate growth and advection \citep{Kessler1969, Wisner_etal1972, Lin_etal1983}.
On the other hand, in the bin method, the size distributions of hydrometeors are discretized into several bins.
The mass (and/or number) of hydrometeors in each bin are prognostic variables \citep{Berry1967, Li_etal2009, Onishi_Takahashi2012}.

With the recent increase in computational power, Lagrangian methods have also been developed \citep{Liu_etal2023}.
\citet{Shima_etal2009} proposed a super-droplet method (SDM) combined with large-eddy simulation (LES).
SDM is a Lagrangian particle tracking-based cloud modeling, in which each computational particle called super-droplet represents multiple real precipitation particles with the same properties.
In this method, the advection of hydrometeors is calculated as the motion of super-droplets.
The collision-coalescence growth is calculated by a stochastic scheme \citep{Shima_etal2009, Shima_etal2020}.

In parallel, direct tracking method combined with direct numerical simulation (DNS) has also been developed \citep{Onishi_etal2015, Saito_Gotoh2018, Chen_etal2018}.
In this approach, the motion and growth of individual precipitation particles are explicitly calculated.
Compared to SDM, direct tracking methods reduce stochastic treatments and can more accurately represent collision-coalescence growth.
In addition, as DNS resolves the smallest scales of turbulence, direct tracking methods enable a more explicit representation of turbulence-particle interactions than SDM and Euler approaches.

The influence of turbulence on cloud microphysics is known to be significant and has been a subject of active discussion for decades \citep{Vaillancourt_Yau2000, Shaw2003, Grabowski_Wang2013}.  
\citet{Saffman_Turner1956} proposed that, during the early stage of cloud formation, turbulence can promote collision-coalescence among small droplets, leading to the formation of larger droplets. 
In their study, the influence of turbulence on the growth of precipitation particles was formulated for the first time.

Since the 1990s, advances in computational power have enabled the use of computationally expensive numerical approaches. 
In fluid engineering, the influence of turbulence on particle motion has been investigated numerically using direct numerical simulation (DNS) of turbulence combined with particle tracking calculations \citep{Squires_Eaton1990, Wang_Maxey1993, Sundaram_Collins1997, Reade_Collins2000}. 
Following this development, similar approaches were introduced in cloud microphysics, where DNS coupled with particle tracking has been used to examine the effects of turbulence on cloud microphysical processes \citep{Vaillancourt_etal2001, Franklin_etal2005, Woittiez_etal2009}. 

Through these DNS-based studies, a theoretical framework for particle collisions that incorporates the effects of turbulence has been established.
\citet{Wang_etal1998} proposed the collision kernel $K^{c}_{1,2}$ between particles 1 and 2 expressed as 
\begin{equation}
  K^{c}_{1,2}
  = 2 \pi R_{1,2}^2 \left\langle \left| w_{r1,2} (r = R_{1,2}) \right| \right\rangle g_{1,2} (r = R_{1,2}),
\end{equation}
where $R_{1,2}$ is the sum of the particle radii (collision radius), 
$\left\langle \left| w_{r1,2} (r = R_{1,2}) \right| \right\rangle$ is the mean radial relative velocity (RRV) at contact, 
and $g_{1,2} (r = R_{1,2})$ is the radial distribution function (RDF) at contact. 
This formulation describes two primary mechanisms by which turbulence enhances particle collisions: 
one is the increase in the RRV due to turbulent shear.
The other is the preferential concentration of inertial particles, which leads to the increase in RDF at contact.

Subsequent studies in cloud microphysics have further advanced DNS with particle tracking through the incorporation of more detailed microphysical processes.
For example, \citet{Michel_etal2023} and \citet{Ababaei_etal2024} developed DNS that account for particle-particle interactions, such as flow disturbances induced by particle motion and lubrication forces. 
In addition, \citet{Onishi_etal2015}, \citet{Saito_Gotoh2018}, and \citet{Chen_etal2018} developed models that incorporate collision-coalescence processes, enabling the direct simulation of precipitation particle growth in turbulent flows. 
This method is particularly valuable for investigating how turbulence affects collision-coalescence growth of droplets, and ultimately precipitation formation.

However, due to their high computational cost, the use of the method has been restricted to fundamental research on cloud microphysics and turbulence.
Most of the previous studies adopted cubic domains with periodic boundaries \citep{Onishi_etal2015, Saito_Gotoh2018, Chen_etal2018}.
Although the periodic box domain is numerically convenient, it is not physically realistic: real atmospheric clouds have vertical structure in terms of, for example, supersaturation ratio or droplet size distribution.
However, the periodic box domain cannot represent such vertical structures of the clouds.

To overcome this limitation, \citet{Kunishima_Onishi2018} introduced a vertically-elongated quasi-1D computational domain, which covers from the ground to the cloud top, to a direct tracking method.
In addition, a model for cloud condensation nuclei (CCN) activation was implemented to the method.
From these extensions, the all warm-cloud microphysical processes (CCN activation, condensation/evaporation growth, collision-coalescence growth, and sedimentation) could be treated by the direct tracking method.
However, they did not incorporate turbulence into their computation.

Thus, the purpose of the present study is to extend the framework proposed in \citet{Kunishima_Onishi2018} and incorporate turbulence.
We aim to investigate how turbulence influences the warm-cloud microphysics using a vertically elongated quasi-one-dimensional domain and implemented turbulence fields.
Our approach is positioned as an intermediate between the conventional DNS studies and other cloud microphysics schemes.
It inherits the advantages of direct tracking methods, which are reduced stochastic treatment and explicit representation of turbulence-microphysics interactions.
It also enables simulations to span the all warm-cloud microphysical processes through the use of a vertically-elongated computational domain.

\section{Numerical simulation}
\label{sec:numerical_simulation}
\subsection{Overview}
\begin{figure}[!tb]
    \centering
    \includegraphics[width=0.6\textwidth]{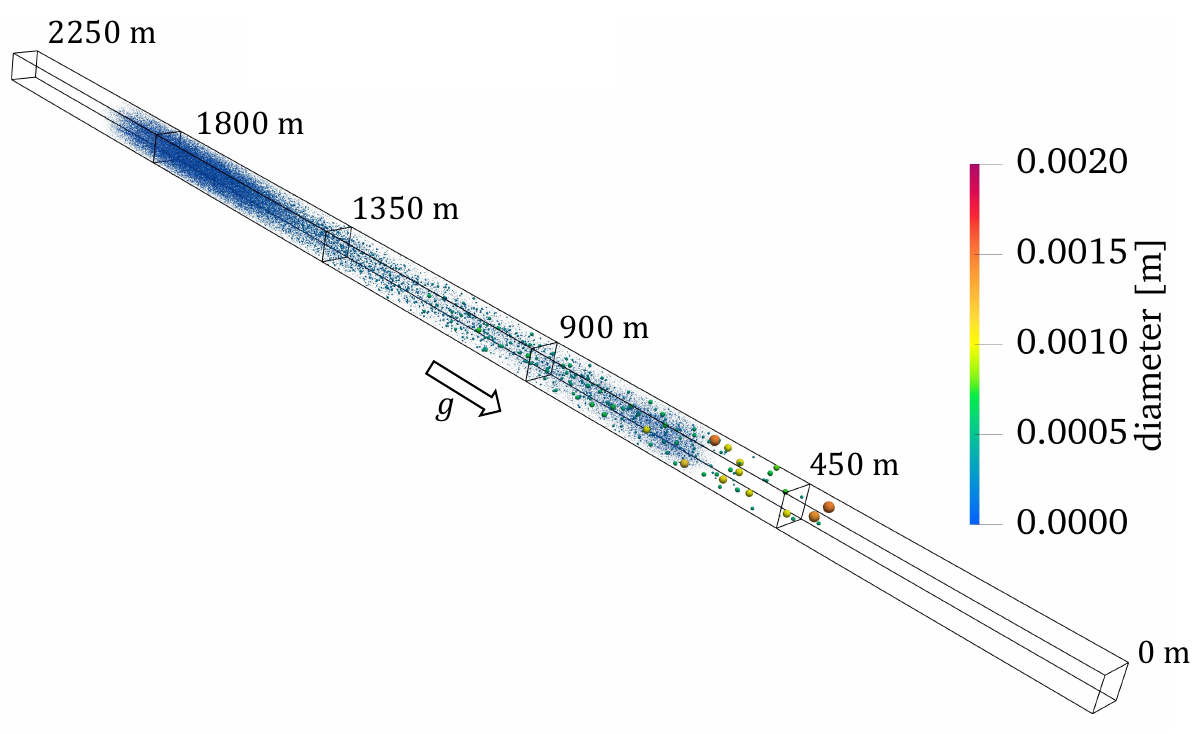}
    \caption{Schematic of vertically-elongated quasi-1D computational domain.
            Liquid droplets at $t=1000\ \mathrm{s}$ are visualized.}
    \label{fig:somen_domain}
\end{figure}
Numerical simulation is based on Lagrangian Cloud Simulator (LCS) \citep{Onishi_etal2015}, which is a DNS-based Lagrangian particle tracking simulation for warm-cloud microphysics.
The simulation consists of a combination of Euler framework for flow and scalar (water vapor and temperature) fields and Lagrangian framework for tracking precipitation particles individually.

LCS with vertically-elongated quasi-1D computational domain is based on \citet{Kunishima_Onishi2018}.
Figure \ref{fig:somen_domain} shows the schematic image of the vertically-elongated domain.
We can consider the all warm-cloud microphysical processes (CCN activation, condensation/evaporation growth, collision-coalescence growth, and sedimentation) using the vertically-elongated domain covering from the ground to the cloud top.

\begin{figure}[!tb]
    \centering
    \includegraphics[width=0.6\textwidth]{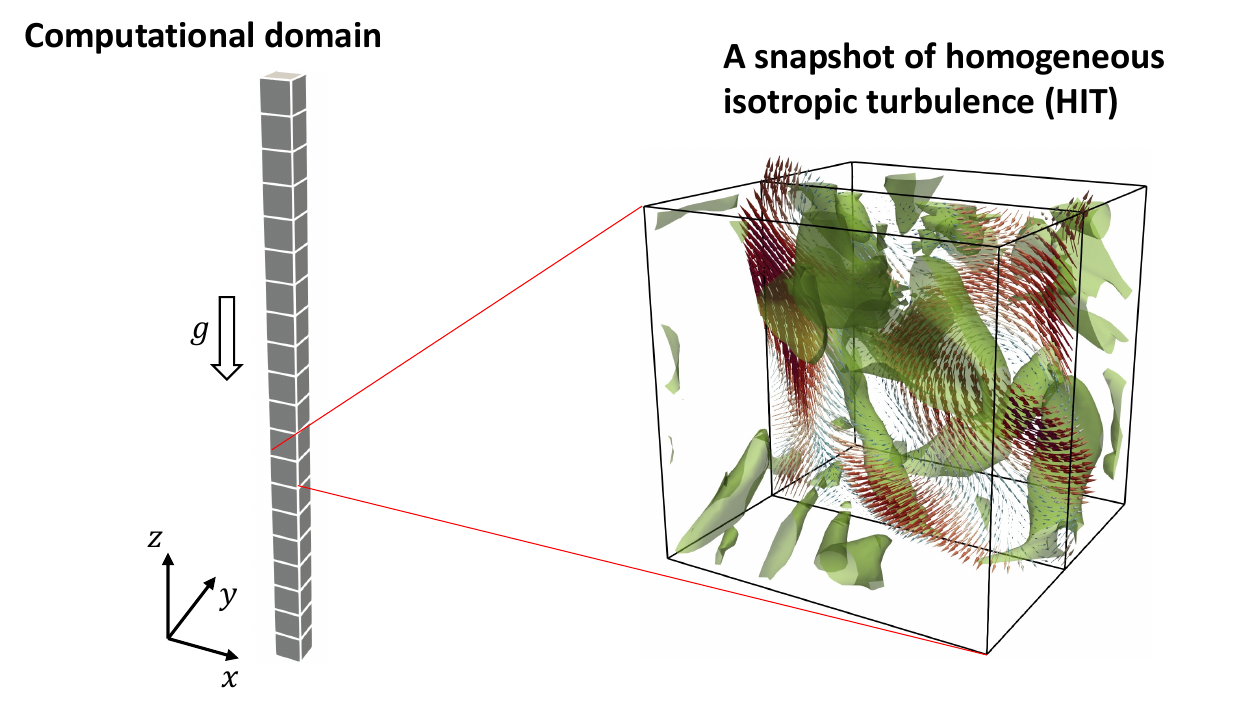}
    \caption{A snapshot of homogeneous isotropic turbulence (HIT) is repeated in the vertical direction to fill the vertically-elongated domain.
            In the figure of HIT, the colored isosurfaces and arrows represent the Q-criterion isosurfaces and the velocity vectors on the $x-z$ plane, respectively.}
    \label{fig:hit_snapshot}
\end{figure}
In this study, a snapshot of homogeneous isotropic turbulence (HIT) is introduced into the vertically-elongated domain to account for turbulence effects on cloud microphysics.
Figure \ref{fig:hit_snapshot} illustrates how the snapshot is implemented.
The snapshot is stacked in the vertically periodic direction of the vertically-elongated domain.

When particles are falling with gravity in a periodic turbulent field, it is known that the particles form curtainlike structures because of the interactions between a large scale of eddy and terminal velocity of particles \citep{Woittiez_etal2009}. 
In order to avoid those artificial structures, we added random perturbations to the horizontal particle positions when falling particles cross the horizontal boundary of the vertically aligned cubic snapshot of HIT.
The perturbations were set to follow a uniform distribution between $-1\ \mathrm{mm}$ and $+1\ \mathrm{mm}$, which was smaller than the average distance between particles ($n_0^{-1/3} = 2.71\ \mathrm{mm}$, where $n_0 = 5 \times 10^7\ \mathrm{m^{-3}}$ is the number density of particles).

\subsection{Euler framework for water vapor fields}
The governing equation for the water vapor mixing ratio $Q_{\upsilon}$ is given by
\begin{equation}
  \label{eq:transport_equation}
  \left( \frac{\partial}{\partial t} + \mathbf{U} \cdot \mathbf{\nabla} \right) Q_{\upsilon} = \kappa_{q} \nabla^2 Q_{\upsilon} + S_{q}, 
\end{equation}
where $\mathbf{U}$ is the flow velocity,
$\kappa_{q}$ is the diffusion coefficient for water vapor in air,
and $S_{q}$ is the source term due to CCN activation and condensation/evaporation growth of droplets.
While the advection of the water vapor field is calculated explicitly, flow and temperature fields are prescribed (see Section \ref{sec:kinematic_model}).

\subsection{Lagrangian framework for precipitation particles}
\subsubsection{Motion of particles}
The governing equation of motion for the $i$th particle is given by
\begin{equation}
  \label{eq:particle_motion}
  \frac{\mathrm{d} \mathbf{V}_{\mathrm{p},i}}{\mathrm{d} t} = - \frac{\alpha_i}{\tau_{\mathrm{p},i}} \left[ \mathbf{V}_{\mathrm{p},i} - \left( \mathbf{U}(\mathbf{x}_{\mathrm{p},i}) + \mathbf{u}(\mathbf{x}_{\mathrm{p},i}) \right) \right] + \mathbf{f}_{\mathrm{coll},i} + \mathbf{g},
\end{equation}
where $\mathbf{V}_{\mathrm{p},i}$ is the velocity of the particle,
$\mathbf{U}(\mathbf{x}_{\mathrm{p},i})$ is the flow velocity at the particle position $\mathbf{x}_{\mathrm{p},i}$,
$\mathbf{u}(\mathbf{x}_{\mathrm{p},i})$ is the velocity perturbation caused by surrounding particles, 
and $\tau_{\mathrm{p},i}$ is the particle relaxation time.
$\alpha_i$ is the non-linear drag coefficient defined as $\alpha_i = 1 + 0.15 Re_{\mathrm{p},i}^{0.687}$, where $Re_{\mathrm{p},i}$ is the particle Reynolds number \citep{Rowe_Henwood1961}.
$\mathbf{f}_{\mathrm{coll},i}$ is the force due to collisions with other particles, and $\mathbf{g}$ is the gravitational acceleration.

\subsubsection{Growth of particles}
CCN activation is treated based on a stochastic model \citep{Twomey1959}.
This model represents the relationship between supersaturation ratio and the number of activated droplets.
The numbers and sizes of newly activated droplets are computed at each time step as a function of the current supersaturation ratio.
Maritime conditions are assumed, with parameter values following \citet{Onishi_Takahashi2012}.
The sizes of newly activated droplets follow an exponential distribution, with the average-mass droplet radius set to $11\ \mathrm{\mu m}$ \citep{Onishi_Takahashi2012}.

The growth rate of particles by condensation/evaporation is calculated by a radius growth equation that depends on the ambient supersaturation ratio and particle size.
The supersaturation ratio at the position of each particle is computed by trilinear interpolation from the surrounding grid points.
Condensation/evaporation growth is coupled with the water vapor field via the term $S_q$ in Eq. (\ref{eq:transport_equation}).

Collision-coalescence growth is computed geometrically.
The trajectories of particle pairs between $t$ and $t+\Delta t$ are assumed to be linear, and are checked individually to determine whether they collide.
When the distance between two particles becomes smaller than the sum of their radii (collision radius), a collision is detected.
The hydrodynamic interaction between particles is considered by solving the Stokes flow around two approaching spheres.
The coalescence efficiency is assumed to be 1 (i.e., colliding droplet pairs are all coalesced.).
The breakup process is not considered, as the droplet sizes in the present simulations remain within a range (Weber number $We \lesssim 1$) where breakup is not significant.
This assumption is consistent with the previous simulations in \citet{Kunishima_Onishi2018}.

\section{Computational settings}
\label{sec:settings}

\subsection{Kinematic model}
\label{sec:kinematic_model}
This study adopted a kinematic model, which prescribes the flow field.
The warm-1 case in the Kinematic Driver (KiD warm-1) \citep{Shipway_Hill2012, Hill_etal2023} was adopted.
This case assumes a simple shallow convective cloud with the prescribed updraft $w_{\mathrm{updraft}}$ defined as
\begin{equation}
  w_{\mathrm{updraft}} \left( t \right) = 
  \begin{cases}
    w_0 \sin \left( \frac{\pi t}{600} \right) & (0\ \mathrm{s} \leq t < 600\ \mathrm{s}) \\
    0                                         & (600\ \mathrm{s} \leq t)
  \end{cases},
\end{equation}
where $w_0 = 2\ \mathrm{m\ s^{-1}}$.
The air parcel is advected by $\int w_{\mathrm{updraft}} \mathrm{d}t = 764\ \mathrm{m}$ during the first $600\ \mathrm{s}$.

The temperature field does not change with time, while the transport of the water vapor field is explicitly computed according to Eq. (\ref{eq:transport_equation}).

\subsection{Initial and boundary conditions}
\label{sec:initial_boundary_conditions}

\begin{table} [!tb]
  \caption{Initial temperature and moisture profiles for KiD warm-1 \citep{Shipway_Hill2012, Hill_etal2023}.}
  \label{tab:kid_warm1_initial}
  \vspace{-5mm}
    \begin{center}
      \begin{tabular}{c | c c}
          \hline
            Altitude           & Potential temperature  & Water vapor mixing ratio  \\
            $z\ [\mathrm{m}]$  & $\Theta\ [\mathrm{K}]$ & $Q_v\ [\mathrm{kg\ kg^{-1}}]$  \\
          \hline
            $2250$             & $306.7$                & $0.00697$  \\
            $740$              & $297.9$                & $0.0138$   \\
            $0$                & $297.9$                & $0.0150$   \\
          \hline
      \end{tabular}
    \end{center}
\end{table}

Table \ref{tab:kid_warm1_initial} shows the initial temperature and moisture profiles for KiD warm-1.
The gradient of the potential temperature is zero below $z = 740\ \mathrm{m}$, so that the air is assumed to be well-mixed and neutrally stratified in this region.
Due to this profile, the cloud base is formed at an altitude of approximately $600 \sim 700\ \mathrm{m}$, and the cloud top reaches around $2000\ \mathrm{m}$. 
In this paper, based on these properties, the lower, middle, and upper layers of the cloud are defined as the altitude ranges of $700-750\ \mathrm{m}$, $1200-1250\ \mathrm{m}$, and $1700-1750\ \mathrm{m}$, respectively.

At the initial time, all particles were cloud condensation nuclei (CCNs), which are dry aerosols to be activated to cloud droplets by CCN activation.
CCNs were uniformly distributed in the domain at the initial time with the number concentration $n_{\mathrm{p}0} = 5 \times 10^7\ \mathrm{m^{-3}}$, which assumes maritime conditions \citep{Onishi_Takahashi2012}.

Periodic boundary conditions were imposed on the particle field in all three directions, except for particles that reach the ground surface (z = 0).
These particles, hereafter referred to as surface-reaching raindrops, are immediately removed from the system.
Due to the prescribed updraft, some particles exit the domain through the top boundary and subsequently reenter from the bottom boundary (ground surface).
Although this treatment may appear unphysical, it does not affect the physical results, as all reentering particles are dry CCN rather than activated liquid droplets.

For the water vapor field, periodic boundary conditions were applied in the horizontal directions.
The top and bottom boundaries for the water vapor field were Neumann boundary conditions with zero gradient, so that the $Q_{\upsilon}$ value near the ground remains at its initial value.

\subsection{Homogeneous isotropic turbulence}
\label{sec:homogeneous_isotropic_turbulence}
A snapshot of homogeneous isotropic turbulence (HIT) was created by LCS \citep{Onishi_etal2015} (flow component only) in a cubic domain with periodic boundary conditions.
For forcing, reduced communication forcing (FDM-RCF) \citep{Onishi_etal2011} was adopted.

\begin{table} [!tb]
\caption{Computational setting for homogeneous isotropic turbulence (HIT).}
\label{tab:hit_setting}
\vspace{-5mm}
\begin{center}
\begin{tabular}{c c c c}
    \hline
        Number of grids   & Representative length & Domain size           & Reynolds number    \\ 
        $N\ [\mathrm{-}]$ & $L_0\ [\mathrm{m}]$   & $L_x\ [\mathrm{m}]$   & $Re\ [\mathrm{-}]$ \\
    \hline
        $32$              & $7.96 \times 10^{-4}$ & $5.00 \times 10^{-3}$ & $1.59$           \\
    \hline
\end{tabular}
\end{center}
\end{table}

\begin{table} [!tb]
\caption{Statistics of homogeneous isotropic turbulence (HIT) obtained in a statistically steady state.}
\label{tab:hit_statistics}
\vspace{-5mm}
\begin{center}
\begin{tabular}{c c c c}
    \hline
      Kolmogorov scale      & Taylor microscale       & Taylor microscale            & Energy  \\ 
                            &                         & Reynolds number              & dissipation rate \\
      $\eta\ [\mathrm{m}]$  & $\lambda\ [\mathrm{m}]$ & $Re_{\lambda}\ [\mathrm{-}]$ & $\varepsilon\ [\mathrm{m^2\ s^{-3}}]$ \\
    \hline
      $4.27 \times 10^{-4}$ & $1.97\times 10^{-3}$    & $5.47$                       & $1.01\times 10^{-1}$ \\
    \hline
\end{tabular}
\end{center}
\end{table}

Table \ref{tab:hit_setting} shows the computational settings of HIT.
The domain size of the HIT was set to be the same as the horizontal size of the vertically-elongated domain.
Table \ref{tab:hit_statistics} shows the statistics of turbulence obtained in a statistically-steady state.
The energy dissipation rate was set to be around the maximum value measured in cumulus clouds \citep{Vaillancourt_Yau2000}.
Due to the reduction of computational cost, the snapshot was averaged over $2^3$ grid points, resulting in a total grid points of $16^3$.
We confirmed that the Taylor microscale Reynolds number $Re_{\lambda}$ and the energy dissipation rate $\varepsilon$ were not changed significantly by this averaging.

In this study, the prescribed flow field $\mathbf{U} = (u, v, w)$ was set to be the sum of the updraft in KiD warm-1 and the HIT flow as
\begin{equation}
  \begin{pmatrix}
    u(\mathbf{x}) \\
    v(\mathbf{x}) \\
    w(\mathbf{x}, t)
  \end{pmatrix}
  =
  \begin{pmatrix}
    u_{\mathrm{HIT}}(\mathbf{x}) \\
    v_{\mathrm{HIT}}(\mathbf{x}) \\
    w_{\mathrm{HIT}}(\mathbf{x}) + w_{\mathrm{updraft}}(t)
  \end{pmatrix},
\end{equation}
where $u_{\mathrm{HIT}}$, $v_{\mathrm{HIT}}$, and $w_{\mathrm{HIT}}$ are the HIT velocity components in the $x$, $y$, and $z$ directions, respectively.

\subsection{Computational domain and case configurations}
\label{sec:computational_domain_and_case_configurations}

\begin{table}[!tb]
  \caption{Computational settings for LAM-case and TURB-case.}
  \label{tab:computational_settings}
  \vspace{-5mm}
  \begin{center}
    \begin{tabular}{c c c c c}
      \hline
                & Domain size                                         & Grid size                                     & Time interval            & Duration Time  \\ 
                & $L_x^2 \times L_z\ [\mathrm{m^3}]$                  & $\Delta x^2 \times \Delta z\ [\mathrm{mm^3}]$ & $\Delta t\ [\mathrm{s}]$ & $T_{\mathrm{total}}\ [\mathrm{s}]$ \\
      \hline
      LAM-case  & $(5.00 \times 10^{-3})^2 \times (2.25 \times 10^3)$ & $1.67^2 \times 9.77$                          & $1.25 \times 10^{-3}$    & $1800$ \\
      TURB-case & $(5.00 \times 10^{-3})^2 \times (2.25 \times 10^3)$ & $0.31^2 \times 0.31$                          & $5.00 \times 10^{-5}$    & $1800$ \\
      \hline
    \end{tabular}
  \end{center}
\end{table}

Two simulation cases were carried out with the case names LAM-case and TURB-case.
LAM stands for laminar (without turbulent velocity fluctuations) and TURB stands for turbulent (with turbulent velocity fluctuations).
The prescribed flow field consists only of the updraft in LAM-case ($\mathbf{U} = (0, 0, w_{\mathrm{updraft}})$), whereas it includes both the updraft and the HIT flow in TURB-case.

Table \ref{tab:computational_settings} shows the settings of the vertically-elongated domain for both cases.
The horizontal size of the domain was determined with consideration of the hydrodynamic interactions (HI) and the computational cost:
The influential length scale of HI is approximately 10 times the particle radius when tha particle Reynolds number is small \citep{Ayala_etal2007, Onishi_etal2013}.
The effect of HI becomes smaller as the particle size increases and is less significant for particles with $r > 40\ \mathrm{\mu m}\ (Re_p > 1)$ \citep{Kunishima_Onishi2018}.
On the other hand, the computational cost highly depends on the number of particles, which is proportional to the horizontal area of the domain.
Thus, the horizontal size of the domain was set to be $5\ \mathrm{mm}$.
For particles with $r \lesssim 40\ \mathrm{\mu m}$, the influential length scale of HI is $\sim 400\ \mathrm{\mu m}$.
The chosen domain size is much larger than the influential length scale of HI, ensuring that hydrodynamic interactions are properly resolved.
The vertical size of the domain was set to be $2250\ \mathrm{m}$, which covers from the ground to the cloud top in the KiD warm-1.

In LAM-case, the grid size and the time interval were the same as those in \citet{Kunishima_Onishi2018}.
In TURB-case, the grid size was set to be the same as that of the snapshot of homogeneous isotropic turbulence (see Section \ref{sec:homogeneous_isotropic_turbulence}).
The time interval in TURB-case was $25$ times smaller than that in LAM-case.
This was determined by the ratio of the vertical grid size between LAM-case and TURB-case ($9.77\ \mathrm{mm} / 0.31\ \mathrm{mm} = 31.5$), so that droplets would not move more than one grid within a single time interval.

\section{Results and Discussion}
\label{sec:results_and_discussion}

\begin{figure}[!tb]
    \centering
    \includegraphics[width=0.9\textwidth]{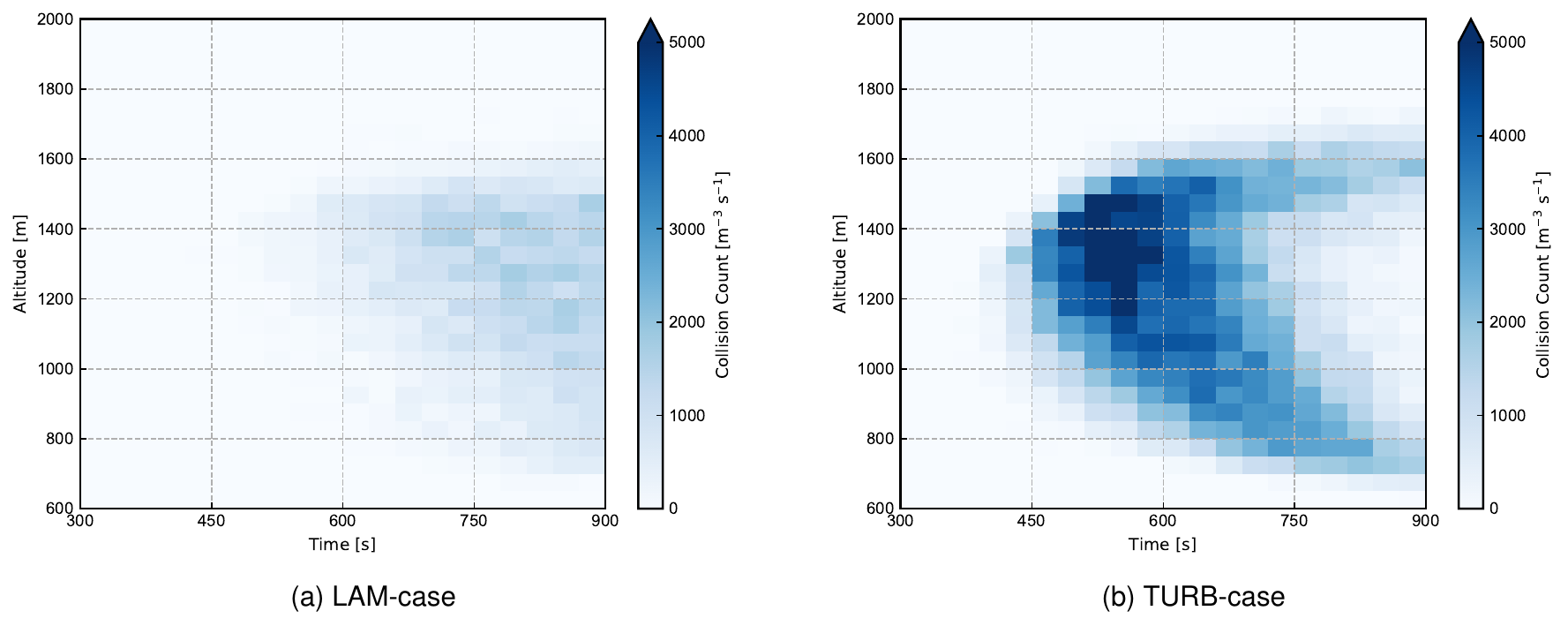}
    \caption{The collision frequency $[\mathrm{m^{-3} s^{-1}}]$ between droplets with $d_{\mathrm{p}} < 80\ \mathrm{\mu m}$ that result in droplets with $d_{\mathrm{p}} \ge 80\ \mathrm{\mu m}$.}
    \label{fig:num_of_coll_80um}
\end{figure}

\begin{figure}[!tb]
  \centering
  \includegraphics[width=0.8\textwidth]{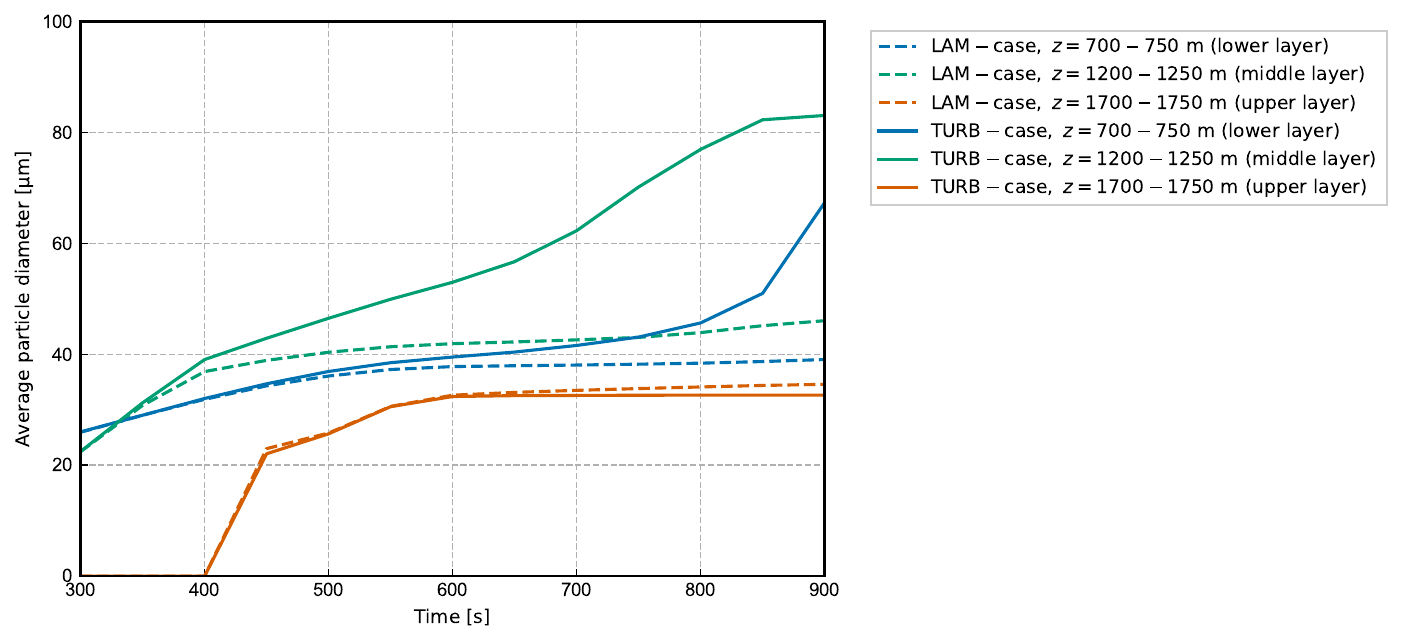}
  \caption{Temporal evolutions of average diameter (mass-weighted average) at three representative altitudes (lower layer: $z=700-750\ \mathrm{m}$, middle layer: $z=1200-1250\ \mathrm{m}$, and upper layer: $z=1700-1750\ \mathrm{m}$).}
  \label{fig:average_diameter_altitude}
\end{figure}

\begin{figure}[!tb]
    \centering
    \includegraphics[width=0.8\textwidth]{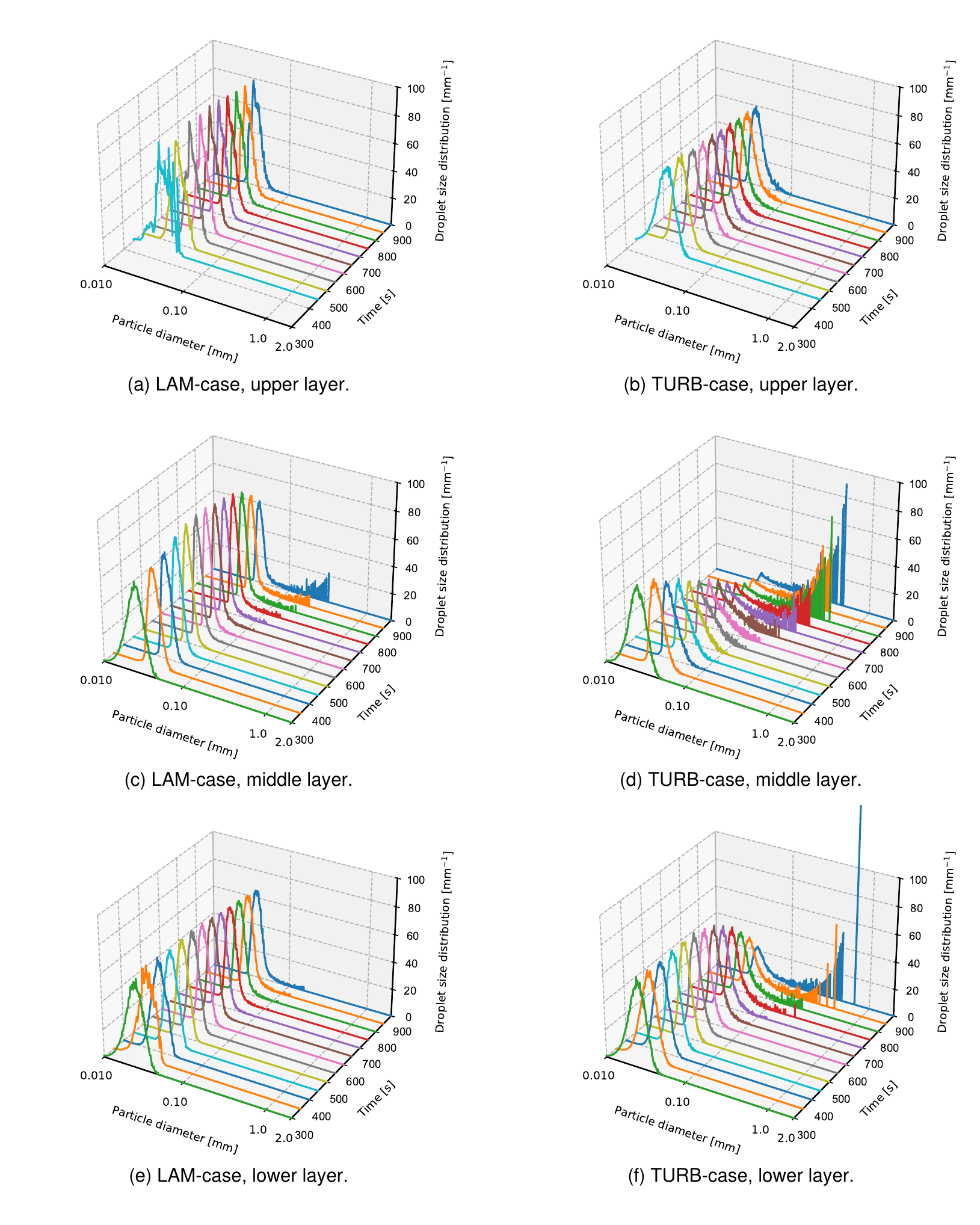}
    \caption{Temporal evolutions of droplet size distribution (DSD) at three altitude ranges in the cloud (lower layer: $z=700-750\ \mathrm{m}$, middle layer: $z=1200-1250\ \mathrm{m}$, and upper layer: $z=1700-1750\ \mathrm{m}$).}
    \label{fig:dsd_time_evo}
\end{figure}

Figure \ref{fig:num_of_coll_80um} shows the heat map of the collision frequency $[\mathrm{m^{-3} s^{-1}}]$ between droplets with $d_{\mathrm{p}} < 80\ \mathrm{\mu m}$ that result in droplets with $d_{\mathrm{p}} \ge 80\ \mathrm{\mu m}$.
In LAM-case, the collision frequency was up to about $2000\ \mathrm{m^{-3}\ s^{-1}}$.
The frequency was smaller than $1000\ \mathrm{m^{-3}\ s^{-1}}$ during the early developing stage ($t < 600\ \mathrm{s}$) in LAM-case.
In contrast, the frequency reaches over $5000\ \mathrm{m^{-3}\ s^{-1}}$ in the middle layer of the cloud during the early developing stage in TURB-case.
These results show that turbulence promoted the collisions of droplet pairs in the middle layer during the early stage.

In addition, the average diameter ratio of the colliding droplets with the threshold diameter of $80\ \mathrm{\mu m}$ was calculated from $300\ \mathrm{s}$ to $600\ \mathrm{s}$.
The average diameter ratio was $1.88$ in LAM-case, while it was $1.66$ in TURB-case.
This result indicates that turbulence promoted the collisions between droplets with similar sizes (autoconversions) during the early stage.

Figure \ref{fig:average_diameter_altitude} shows the temporal evolution of average diameter at three representative altitudes in the cloud (lower layer: $z=700-750\ \mathrm{m}$, middle layer: $z=1200-1250\ \mathrm{m}$, and upper layer: $z=1700-1750\ \mathrm{m}$).
In LAM-case, the average diameters at the three altitudes exhibit little difference.
In contrast, the time evolution of the average diameter shows a clear difference among the three altitudes in TURB-case.
First, a discrepancy from LAM-case is observed in the middle layer after around $400\ \mathrm{s}$.
Later, at around $800\ \mathrm{s}$, a sharp increase is observed in the lower layer.

These results in TURB-case can be explained by the difference in scenario of droplet growth in each layer.
In the upper layer, condensation occurred later than in the lower layers.
Therefore, droplet growth was delayed and the average diameter remained small during the early stage.
In the lower layer, droplets started to form at an early time. 
However, these newly formed droplets were transported upward by the updraft.
As a result, the average diameter did not increase significantly at the early developing stage.
In the middle layer, droplets formed in the lower layer were transported upward while undergoing autoconversion by turbulence.
Consequently, the average diameter increased earlier than in the lower and upper layers, which is shown around $400\ \mathrm{s}$.

After the early developing stage, droplets fell due to the gravity.
In the middle layer, some growing droplets fell from higher altitudes, whereas others continued to fall toward lower altitudes.
As a result, the average diameter increased gradually.
In the lower layer, large droplets formed at higher altitudes, where collision-coalescence growth was enhanced by turbulence during the early stage.
These droplets then fell into this region while growing through the collection of smaller droplets (accretion).
This led to a rapid increase in the average diameter at around $800\ \mathrm{s}$.

Figure \ref{fig:dsd_time_evo} shows the temporal evolutions of the droplet size distribution (DSD) $f \left( d_{\mathrm{p}} \right)\ \left[ \mathrm{mm^{-1}} \right]$ at three representative altitudes in the cloud (lower: $z=700-750\ \mathrm{m}$, middle: $z=1200-1250\ \mathrm{m}$, and upper: $z=1700-1750\ \mathrm{m}$ layers).
The DSD was normalized by total water mass, so that the integral over the entire diameter range is unity ($\int_0^\infty f(d_{\mathrm{p}}) \, \mathrm{d} d_{\mathrm{p}} = 1$).

To characterize the broadening of the droplet size distribution (DSD), we focus on the time at which droplets with diameters exceeding $d_{\mathrm{p}} = 100\ \mathrm{\mu m}$ first appear.
Such large droplets do not appear until after approximately $600\ \mathrm{s}$ at all altitudes in LAM-case.
In contrast, droplets larger than $100\ \mathrm{\mu m}$ emerge much earlier in the middle layer, at around $400 \sim 450\ \mathrm{s}$ in TURB-case.
This early broadening of the DSD in the middle layer clearly demonstrates that turbulence promotes autoconversions during the early developing stage.
Following this, large droplets appear in the lower layer after approximately $650\ \mathrm{s}$ in TURB-case.
This demonstrates the rapid growth in the lower layer after the early stage.
These trends are consistent with those observed in the temporal evolution of the average diameter (Fig. \ref{fig:average_diameter_altitude}).

A comparison of the DSD between the middle and lower layers in TURB-case further highlights the difference in growth scenarios.
When comparing the middle layer to the lower layer in TURB-case, the average diameter at $t=700\ \mathrm{s}$ in the middle layer is almost the same as that at $t=900\ \mathrm{s}$ in the lower layer.
However, the DSD at $t=700\ \mathrm{s}$ in the middle layer is clearly different from that at $t=900\ \mathrm{s}$ in the lower layer.
This also indicates the discrepancy in the growth scenario between the middle and lower layers in TURB-case.

\begin{figure}[!tb]
    \centering
    \includegraphics[width=0.9\textwidth]{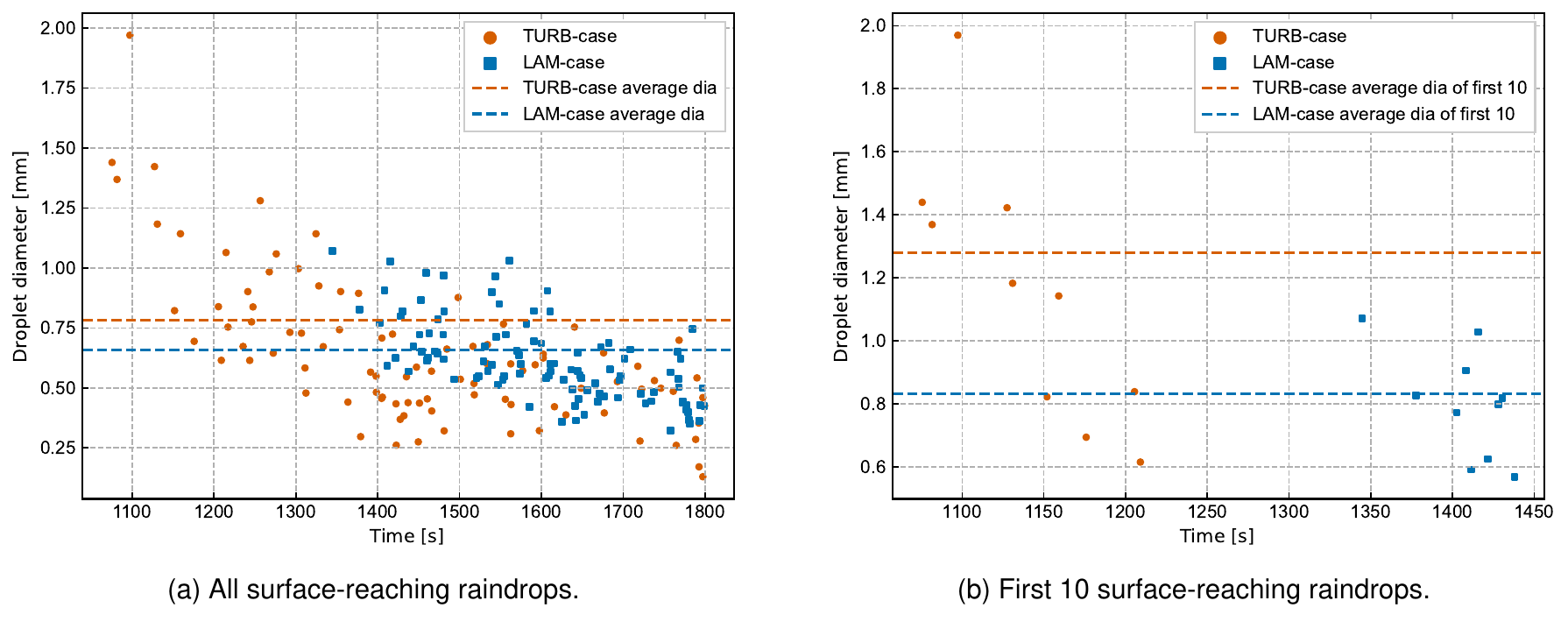}
    \caption{The diameter of surface-reaching raindrops and the time when they reach the ground.}
    \label{fig:dia_raindrops_time}
\end{figure}

Figure \ref{fig:dia_raindrops_time} shows the relationship between the diameter of surface-reaching raindrops and the time when they reach the ground.
Figure \ref{fig:dia_raindrops_time} (a) presents the data for all surface-reaching raindrops, while Figure \ref{fig:dia_raindrops_time} (b) focuses on the first 10 surface-reaching raindrops.
The surface precipitation onset occurs approximately $270\ \mathrm{s}$ earlier in TURB-case than in LAM-case.
Furthermore, Fig. \ref{fig:dia_raindrops_time} (a) shows that, the size of surface-reaching raindrops is larger in TURB-case by a factor of 1.2 compared to LAM-case.
This effect is even more pronounced for the first 10 surface-reaching raindrops (Fig. \ref{fig:dia_raindrops_time} (b)), with sizes larger by a factor of 1.5.
These results indicate that promoted collisions caused earlier precipitation onset and larger initial surface-reaching raindrops with the presence of turbulence.

\begin{figure}[!tb]
  \centering
  \includegraphics[width=0.65\textwidth]{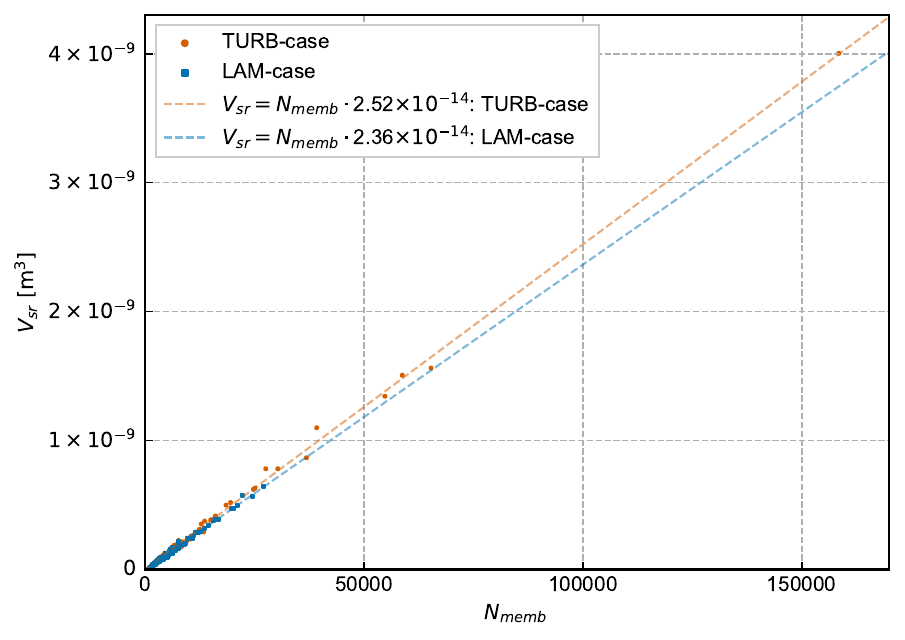}
  \caption{The volume of surface-reaching raindrops $V_\mathrm{sr}\ [\mathrm{m}^3]$ against the number of constituent particles $N_\mathrm{memb}$.}
  \label{fig:surface_precip_volume_vs_number_of_constituent_particles}
\end{figure}

In the Lagrangian tracking simulation, the growth history of each precipitation particle is individually tracked.
Figure \ref{fig:surface_precip_volume_vs_number_of_constituent_particles} shows the relationship between the volume of surface-reaching raindrops, $V_\mathrm{sr}\ [\mathrm{m}^3]$, and the number of constituent particles forming each raindrop, $N_\mathrm{memb}$.
Here, the constituent particle is defined as the single activated droplet before any collisions occur.
The slope of the regression line corresponds to the average volume of individual constituent particles, defined as $v_c = V_{\mathrm{sr}} / N_{\mathrm{memb}}$.
The slopes indicate that $v_{c,\mathrm{TURB}} = 2.52 \times 10^{-14}$ (corresponding to $d = 36.4\ \mathrm{\mu m}$ water droplet) and $v_{c,\mathrm{LAM}} = 2.36 \times 10^{-14}$ (correpsonding to $d = 35.6\ \mathrm{\mu m}$ water droplet), yielding a ratio of $v_{c,\mathrm{TURB}} / v_{c,\mathrm{LAM}} = 1.1$.
The factor of 1.1 is non-negligible and will be discussed below.

The numbers of activated droplets during the first $600\ \mathrm{s}$ were nearly the same in the two cases: $N_{\mathrm{act},\mathrm{LAM}} = 1.49 \times 10^6$ and $N_{\mathrm{act},\mathrm{TURB}} = 1.51 \times 10^6$.
In contrast, the total numbers of collisions during the first $600\ \mathrm{s}$ differed substantially due to turbulent enhancement: $n_{\mathrm{coll},\mathrm{LAM}} = 1.72 \times 10^5$ and $n_{\mathrm{coll},\mathrm{TURB}} = 5.47 \times 10^5$.
This difference results in the smaller number of droplets in TURB-case than in LAM-case at $t = 600\ \mathrm{s}$.
The total volume of condensed water was controlled by the KiD warm-1 configuration, and was nearly the same in both cases.
Therefore, the smaller number of droplets in TURB-case led to a larger average volume per droplet than in LAM-case.
Simple estimates show that the ratio of the average volume per droplet in TURB-case, $V_{\mathrm{TURB}}$, was $V_{\mathrm{TURB}} = 1.37 \times V_{\mathrm{LAM}}$.
The factor of $1.37$, originated from the enhanced collisions in TURB case, is different from that of 1.1 obtained for $v_c$.
This discrepancy can be attributed the differences in condensation growth histories between the two cases.
Since condensation growth depends on the droplet size, changes in the droplet size distribution can alter the growth history.
These results confirmed that turbulence affected the condensation growth through the promotion of the collision-coalescence process.

\conclusions  
\label{sec:conclusions}
We developed a new explicit cloud microphysical model that incorporates turbulence using direct numerical simulation (DNS) coupled with a Lagrangian particle tracking method.
The model explicitly accounts for the effects of turbulence on warm-cloud microphysical processes by introducing a snapshot of homogeneous isotropic turbulence into a vertically-elongated quasi-1D computational domain.
This framework enables us to quantify the impact of turbulence on the warm-cloud microphysical processes, including CCN activation, condensation/evaporation growth, collision-coalescence growth, and sedimentation.

Two different cases of simulations, LAM-case (laminar case without turbulent wind fluctuations) and TURB-case (turbulent case with turbulent wind fluctuations), were conducted to elucidate the impact of turbulence on the precipitation processes.
We obtained new insights into the altitude- and time-dependent microphysical statistics of precipitation particles in turbulence, which cannot be captured through conventional DNS researches for a periodic cubic box domain.

Compared to LAM-case, results of TURB-case showed an increase in collisions between droplets with similar sizes in the middle layer during the early developing stage with non-zero vertical velocity ($t < 600\ \mathrm{s}$).
This indicates that turbulence promoted collisions between droplets with similar sizes, namely autoconversion, in the middle layer during the early stage.
After the early stage ($t \ge 600\ \mathrm{s}$), a sharp increase in the average diameter was observed in the lower layer in TURB-case.
These results clearly demonstrate that turbulence promoted autoconversions in the early developing stage and accelerated accretion afterward.

The relationship between the diameter of surface-reaching raindrops and the time when they reach the ground was examined.
The onset of surface precipitation in TURB-case occurred earlier than that in LAM-case.
Moreover, the diameters of the first surface-reaching raindrops were significantly larger in TURB-case than in LAM-case.
These results suggest that turbulence advances the onset of precipitation and increases the size of the first raindrops reaching the surface.

Taking advantage of the Lagrangian particle tracking approach, the collision histories of surface-reaching raindrops were analyzed.
The relationship between the number of constituent particles and the volume of surface-reaching raindrops was investigated.
The constituent particle was defined as the single activated droplet before experiencing any collisions.
The average volume of constituent particles in TURB-case was larger than that in LAM-case.
The volume difference is attributed to the difference in the condensation growth due to the DSD difference.
This result suggests that turbulence affected not only the collision-coalescence growth but also the condensation growth through the promotion of the collision-coalescence process in the early developing stage.



\dataavailability{Access to the simulation results can be granted upon request under a collaborative research framework.} 













\authorcontribution{
M.I. developed the model, performed the simulations, and analyzed the data. M.I. wrote the manuscript. R.O. conceived the study. M.I. and R.O. discussed the results and reviewed the manuscript.
} 

\competinginterests{
The authors declare that they have no competing interests.
} 


\begin{acknowledgements}
The numerical simulations were performed on the supercomputer, TSUBAME 4.0 at Institute of Science Tokyo.
The authors acknowledge the use of ChatGPT (OpenAI) as a language assistance tool to improve the clarity and readability of the manuscript.
The use of this tool was limited to grammar correction and stylistic refinement.
The authors take full responsibility for the content of the manuscript.
\end{acknowledgements}

\bibliographystyle{copernicus}
\bibliography{reference.bib}

\end{document}